\documentclass[conference]{IEEEtran}

\usepackage{cite}
\usepackage{easyReview}
\usepackage{graphicx}
\usepackage{ctable}
\usepackage{xcolor,colortbl}

\newcommand{\squeezeupupupi}{\vspace{-3mm}}

\usepackage{array}
\usepackage{multirow}
\usepackage{comment}
\usepackage{caption}
\usepackage{subfigure}
\usepackage{stfloats}
\usepackage{balance}
\usepackage{float}
\newfloat{footnote}{hb}

\begin{document}
%
\title{Analog Image Denoising with an Adaptive Memristive Crossbar Network}


\author{
\IEEEauthorblockN{O. Krestinskaya $^{1}$, K.N. Salama $^{1}$ and A.P. James$^{2}$ }
\IEEEauthorblockA{$^{1}$King Abdullah University of Science and Technology, Saudi Arabia; $^{2}$Digital University Kerala}}

\maketitle


\begin{abstract}


Noise in image sensors led to the development of a whole range of denoising filters. A noisy image can become hard to recognize and often require several types of post-processing compensation circuits. This paper proposes an adaptive denoising system implemented using analog in-memory neural computing network.
The proposed method can learn new noises and can be integrated into or alone with CMOS image sensors.  
Three denoising network configurations are implemented, namely, (1) single layer network, (2) convolution network, and (3) fusion network.
The single layer network shows the processing time, energy consumption and on-chip area of $3.2\mu \textnormal{s}$, $21n\textnormal{J}$ per image and $0.3\textnormal{mm}^2$ respectively, meanwhile, convolution denoising network correspondingly shows $72m \textnormal{s}$, $236\mu\textnormal{J}$ and $0.48\textnormal{mm}^2$. 
Among all the implemented networks, 
it is observed that performance metrics SSIM, MSE and PSNR show a maximum improvement of $3.61$, $21.7$ and $7.7$ times respectively.

\end{abstract}



%
\IEEEpeerreviewmaketitle

\vspace{-0.25cm}
\section{Introduction}
\vspace{-0.15cm}
The CMOS image sensors capture the pixel information through photo-diodes and CMOS-based amplification circuits \cite{el2005cmos}. The image pixels' noise gets injected due to non-idealities in the integrated devices and is impacted by temperature, frequency of usage, and device parasitics \cite{gow2007comprehensive}. 
Traditionally, 
these analog signal noises are suppressed by dedicated filters \cite{oike2012cmos} and converted to digital domain, which rejects certain amount of signal noise and allows using the data for further noise compensation by a digital microprocessor.
An alternative school of thought is to make the analog sensor intelligent by incorporating denoising filters directly into the pixels \cite{massari2005cmos,chen2019processing}.  
We propose incorporating denoising into pixel sensors using a continuously trainable memristor-based neural network in analog domain for near-sensor on edge processing. The neural network can be trained for different types of noises and compensates for noises originating from the sensor and related interface circuits. 




\begin{figure}[!ht]
    \centering
    \includegraphics[width=88mm]{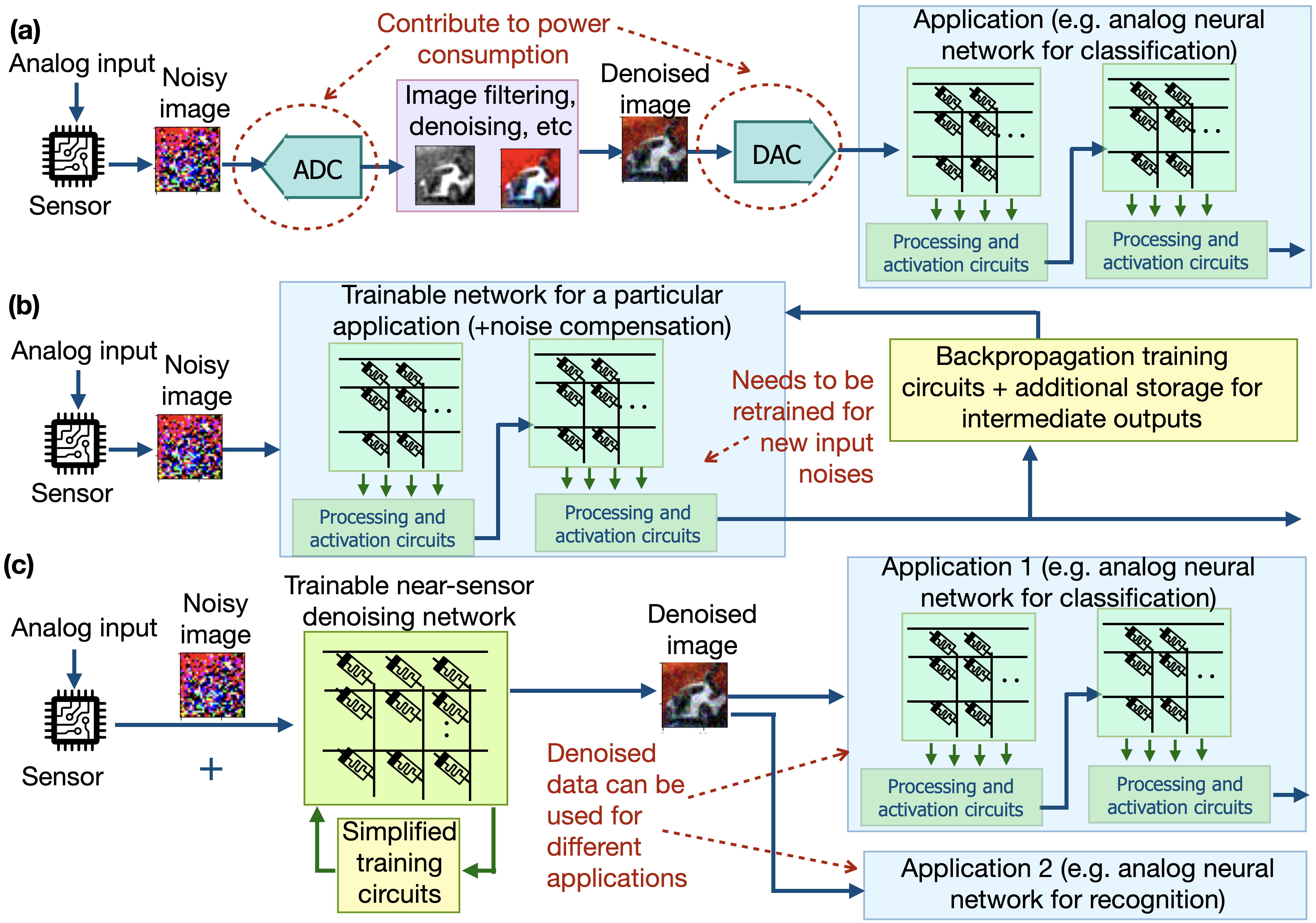}
    \caption{\footnotesize (a) Conversion of analog sensor data to digital domain  with ADC for processing. (b) Retraining neural network for new input noise involving complex backpropagation circuits. (c) Proposed approach using a trainable network layer integrated close to the sensor with reduced training complexity for noise adaptation.}
    \label{f0}
\end{figure}

When using a neural network for inference tasks with noisy images, the commonly known approaches are to apply pre-processing techniques or to incorporate the denoising as part of learning. The system pre-trained with noiseless data fails to process, recognize or classify noisy images \cite{nazare2017deep, diamond2017dirty}.
Most commonly, image sensor noise is compensated by converting sensor outputs to the digital domain using ADC, and applying denoising and filtering techniques \cite{oike2012cmos,kawahito2008cmos} (shown in Fig. \ref{f0} (a)). To use this digital data in an analog memristor-based neural network, it should be converted back to analog domain for dot product computation. In this conversion, ADC and DAC contribute to power consumption, speed and on-chip area overhead \cite{xu2020macsen}. 

For analog memristor-based neural network applications, the desirable approach is reading sensor output directly without converting it to the digital domain (Fig. \ref{f0} (b)) \cite{krestinskaya2020memristive}. However, retraining of the whole network for new noisy inputs is computationally expensive, and requires additional backpropagation circuits and memory contributing to power and area overhead, especially in analog domain \cite{krestinskaya2018analog,krestinskaya2018learning}. Moreover, in a real-time system, the input noise can be unknown, making it challenging to pre-train the system for a particular noise type.

To solve these issues, we propose integrating a small trainable memristor-based denoising network close to the sensor that can be trained with varying noise and is able to denoise the data without retraining the whole system (Fig. \ref{f0} (c)). 

The denoising network can be as small as a single-layer dense network without needing backpropagation circuits to retrain it. In turn, training circuits of reduced complexity are more suitable for implementation on edge devices. The other approach is to introduce small convolution/deconvolution (CNN) network, which requires fewer memristors to store the weights, but has more complex training circuits.

Previously, complex software-based deep neural networks have been used for image denoising \cite{wang2021multi}, \cite{wang2021channel}, \cite{wei20203}. Also, RRAM crossbar based networks  have been useful for image storage \cite{zheng2018error} and edge detection \cite{tang2020fully}. The studies of memristor based neural networks for denoising applications are very limited. 
The existing state-of-the-art works focus on cellular neural networks \cite{zhang2021quantized,slavova2019memristor}, only Gaussian noise \cite{tang2020multilayer,slavova2019memristor,zhang2021quantized} and binary images \cite{tang2020multilayer,zhang2021quantized}, containing limited evaluation of denoising metrics. In this work, we demonstrate the generalized architecture suitable for various types of noises for binary and RGB images, compare the performance to the most common denoising techniques and illustrate the network fusion approach for efficient processing of different types of noises. We evaluate the performance of the proposed system considering memristor non-idealities, and estimate the processing speed, energy consumption and on-chip area of the complete system.

\begin{figure*}[!ht]
    \centering        
   \includegraphics[width=180mm]{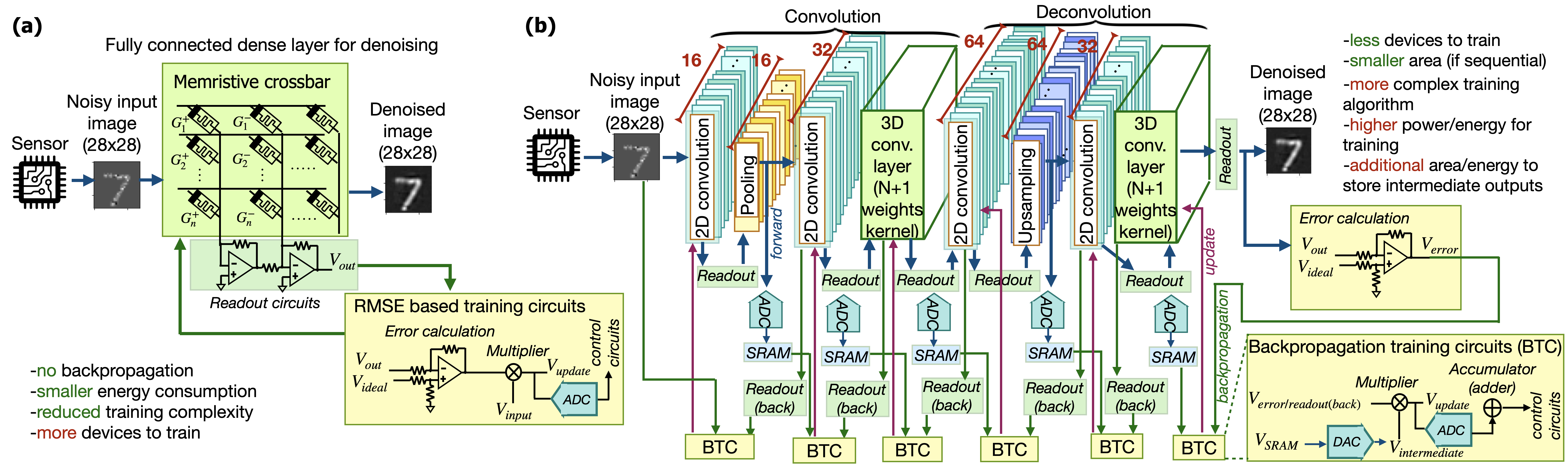}
    \caption{\footnotesize  (a) Single denoising network layer with RMSE-based training. (b) CNN denoising network with backpropagation training.}
    \label{f01}
    \vspace{-0.6cm}
\end{figure*}

\vspace{-0.15cm}
\section{Denoising network for near-sensor processing}
\vspace{-0.1cm}
\subsection{Origin of noise}
\vspace{-0.1cm}
When developing denoising method for image sensors, multiple noise types need to be considered. The most common four types are considered in the paper, and relevance of such noise in image sensors is listed as:
\subsubsection{Gaussian} This noise is additive to the pixel, reflective of Johnson–Nyquist noise (thermal noise) and reset noise of capacitors (kTC noise) \cite{tian2000noise}. The amplifier noise is unavoidable and forms part of the read noise from the image sensors.


\subsubsection{Salt and pepper (S\&P)} This noise originates in data conversion and transmission processes, such as from errors in ADC circuits. It changes a dark pixel to a bright or from bright to the dark. Pixel interpolations and mean filtering are popular approaches to compensate for this noise \cite{cao2011chip}.

\subsubsection{Poisson (Shot) noise} This noise is approximated with a Poisson distribution and at very high-intensity levels as Gaussian noise. It originates from statistical quantum fluctuations that occur in photodetector at high exposures and can also arise from dark leakage current in the image sensor \cite{tian2000noise}.

\subsubsection{Speckle} This noise reflects the unwanted modifications of the desired signal. As the scatterers are not identical, the detected signal becomes sensitive to small changes in object surfaces. This is granular interference to the signal, and is very common in radars and medical images \cite{singh2016speckle}.
\vspace{-0.05cm}



\begin{figure}[!h]
    \centering        
    \includegraphics[width=88mm]{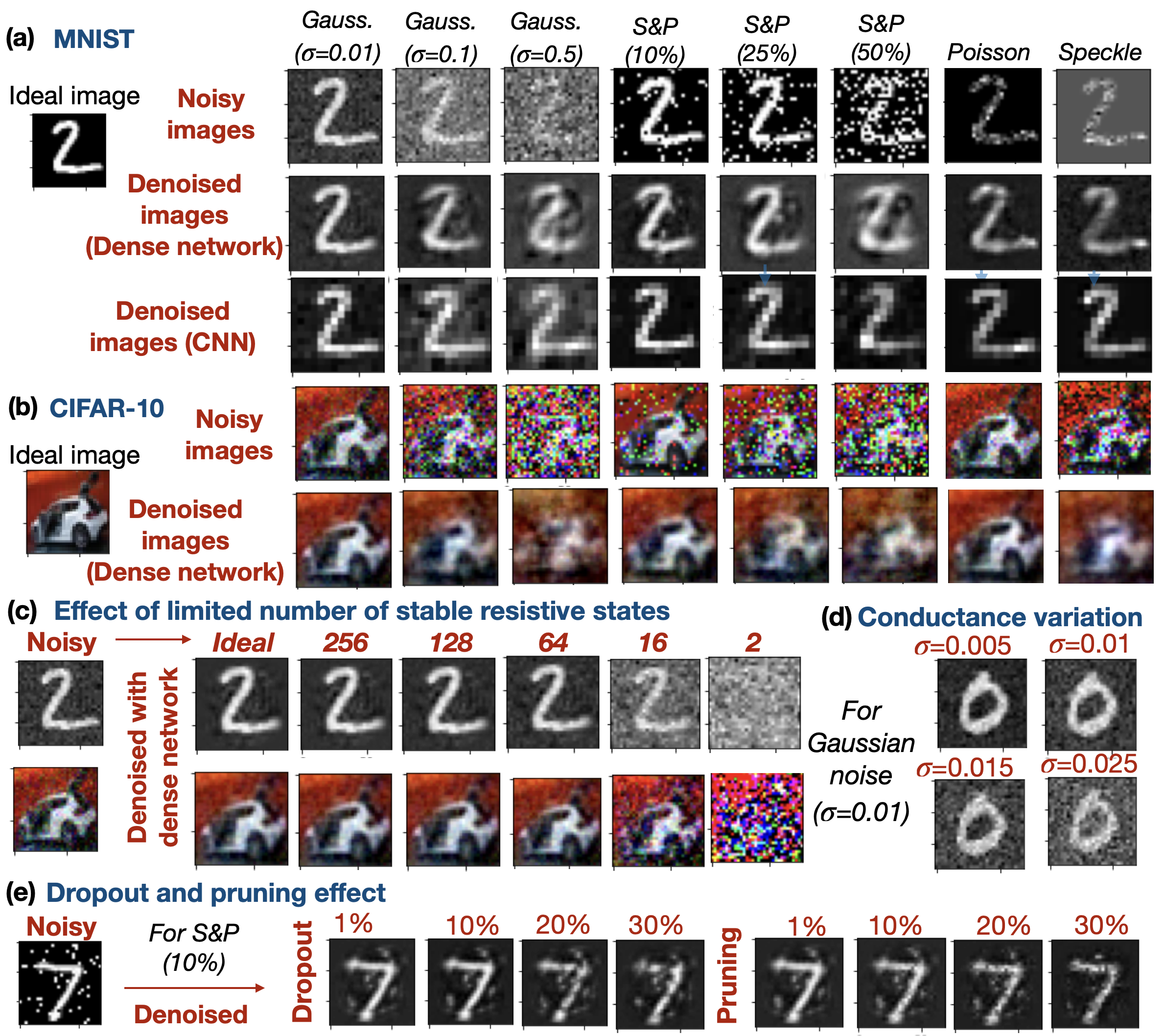}
    \caption{\footnotesize Denoising scores of (a) MNIST images denoised by dense network and CNN, and (b) CIFAR images denoised dense network.  Effect of (c) limited number of stable resistive states and (d) conductance variation in memristors on the denoised images. (e) Effects of the dropout and random pruning.}
    \label{newfigure}
\end{figure}

\begin{table*}[]
\centering
\caption{\small Comparison of SSIM, MSE and PSNR performance metrics for noisy and denoised images, and effect of memristor non-idealities.}
\scriptsize
\begin{tabular}{|lcccccccc|}
\hline
\multicolumn{1}{|c|}{\multirow{2}{*}{\textbf{Noise}}} & \multicolumn{3}{c|}{\textbf{Gaussian}}                                                                                            & \multicolumn{3}{c|}{\textbf{Salt and Pepper}}                                                                                                                             & \multicolumn{1}{c|}{\multirow{2}{*}{\textbf{Poisson}}}                                              & \multirow{2}{*}{\textbf{Speckle}}                                              \\ \cline{2-7}
\multicolumn{1}{|c|}{}                                & \multicolumn{1}{c|}{$\sigma=$0.01}                 & \multicolumn{1}{c|}{$\sigma=$0.1}                  & \multicolumn{1}{c|}{$\sigma=$0.5}                  & \multicolumn{1}{c|}{10\%}                 & \multicolumn{1}{c|}{25\%}                                     & \multicolumn{1}{c|}{50\%}                                     & \multicolumn{1}{c|}{}                                                                               &                                                                             \\ \hline
\multicolumn{9}{|c|}{ \cellcolor{green!10}    \textbf{Comparison of noisy and denoised images, MNIST: {[}Noisy image,  Denoised with dense network,  Denoised with CNN{]}}}                                                                                                                                                                                                                                                                                                                                                                                                                                                      \\ \hline
\multicolumn{1}{|l|}{\textbf{SSIM}}                   & \multicolumn{1}{c|}{{[}0.72,\textbf{0.80},0.74{]}} & \multicolumn{1}{c|}{{[}0.49,\textbf{0.67},0.51{]}} & \multicolumn{1}{c|}{{[}0.23,\textbf{0.53},0.42{]}} & \multicolumn{1}{c|}{{[}0.61,\textbf{0.72},0.61{]}} & \multicolumn{1}{c|}{{[}0.45,\textbf{0.64},0.43{]}}                     & \multicolumn{1}{c|}{{[}0.32,\textbf{0.57},0.34{]}}                     & \multicolumn{1}{c|}{{[}0.83,\textbf{0.71},\textbf{0.71}{]}}                                                           & {[}0.69,\textbf{0.75},0.66{]}                                                           \\ \hline
\multicolumn{1}{|l|}{\textbf{MSE (x10)}}              & \multicolumn{1}{c|}{{[}0.09,\textbf{0.03},0.18{]}} & \multicolumn{1}{c|}{{[}0.99,\textbf{0.14},0.26{]}} & \multicolumn{1}{c|}{{[}4.99,\textbf{0.30},0.43{]}} & \multicolumn{1}{c|}{{[}0.44,\textbf{0.09},0.19{]}} & \multicolumn{1}{c|}{{[}1.0,\textbf{0.16},0.23{]}}                      & \multicolumn{1}{c|}{{[}1.71,\textbf{0.25},0.33{]}}                     & \multicolumn{1}{c|}{{[}0.51,\textbf{0.13},0.22{]}}                                                           & {[}1.14,\textbf{0.18},0.22{]}                                                           \\ \hline
\multicolumn{1}{|l|}{\textbf{PSNR (dB)}}              & \multicolumn{1}{c|}{{[}20.0,\textbf{24.3},17.5{]}} & \multicolumn{1}{c|}{{[}10.0,\textbf{18.5},15.8{]}} & \multicolumn{1}{c|}{{[}3.01,\textbf{15.3},13.7{]}} & \multicolumn{1}{c|}{{[}-, \textbf{20.2}, 17.2{]}}  & \multicolumn{1}{c|}{{[}-, \textbf{17.9}, 16.4{]}}                      & \multicolumn{1}{c|}{{[}-, \textbf{16.1}, 14.9{]}}                      & \multicolumn{1}{c|}{{[}13.1,\textbf{18.9},16.7{]}}                                                           & {[}9.75,\textbf{17.2},16.7{]}                                                           \\ \hline
\multicolumn{9}{|c|}{ \cellcolor{green!10}    \textbf{Comparison of noisy and denoised images, CIFAR: {[}Noisy image,  Denoised with dense network{]}}}                                                                                                                                                                                                                                                                                                                                                                                                                                                                         \\ \hline
\multicolumn{1}{|l|}{\textbf{SSIM}}                   & \multicolumn{1}{c|}{{[}0.67, \textbf{0.76}{]}}     & \multicolumn{1}{c|}{{[}0.28, \textbf{0.61}{]}}     & \multicolumn{1}{c|}{{[}0.11, \textbf{0.40}{]}}     & \multicolumn{1}{c|}{{[}0.49, \textbf{0.70}{]}}     & \multicolumn{1}{c|}{{[}0.29, \textbf{0.60}{]}}                         & \multicolumn{1}{c|}{{[}0.16, \textbf{0.49}{]}}                         & \multicolumn{1}{c|}{{[}0.60, \textbf{0.74}{]}}                                                               & {[}0.24, \textbf{0.51}{]}                                                               \\ \hline
\multicolumn{1}{|l|}{\textbf{MSE (x10)}}              & \multicolumn{1}{c|}{{[}0.10, \textbf{0.05}{]}}     & \multicolumn{1}{c|}{{[}1.00, \textbf{0.10}{]}}     & \multicolumn{1}{c|}{{[}5.00, \textbf{0.23}{]}}     & \multicolumn{1}{c|}{{[}0.29, \textbf{0.07}{]}}     & \multicolumn{1}{c|}{{[}0.82, \textbf{0.69}{]}}                         & \multicolumn{1}{c|}{{[}1.22, \textbf{0.16}{]}}                         & \multicolumn{1}{c|}{{[}0.18, \textbf{0.06}{]}}                                                               & {[}2.90, \textbf{0.16}{]}                                                               \\ \hline
\multicolumn{1}{|l|}{\textbf{PSNR (dB)}}              & \multicolumn{1}{c|}{{[}18.9, \textbf{21.9}{]}}     & \multicolumn{1}{c|}{{[}8.98, \textbf{18.7}{]}}     & \multicolumn{1}{c|}{{[}1.99, \textbf{15.4}{]}}     & \multicolumn{1}{c|}{{[}-, \textbf{20.5}{]}}        & \multicolumn{1}{c|}{{[}-, \textbf{18.6}{]}}                            & \multicolumn{1}{c|}{{[}-, \textbf{17.0}{]}}                            & \multicolumn{1}{c|}{{[}16.4, \textbf{21.3}{]}}                                                               & {[}4.82, \textbf{17.1}{]}                                                               \\ \hline
\multicolumn{9}{|c|}{ \cellcolor{green!10}   \textbf{Effect of memristor non-idealities}}                                                                                                                                                                                                                                                                                                                                                                                                                                                                                            \\ \hline
\multicolumn{1}{|l|}{\multirow{2}{*}{}}               & \multicolumn{4}{c|}{\textbf{\begin{tabular}[c]{@{}c@{}}Limited number of stable resistive states \\ {[}256, 128, 64, 16, 2{]}\end{tabular}}}                                                   & \multicolumn{2}{c|}{\textbf{\begin{tabular}[c]{@{}c@{}}Conductance variation\\ {[}$\sigma =$0.005, 0.01, 0.015, 0.025{]}\end{tabular}}} & \multicolumn{1}{c|}{\textbf{\begin{tabular}[c]{@{}c@{}}Dropout \\ {[}10, 20, 30{]}\%\end{tabular}}} & \textbf{\begin{tabular}[c]{@{}c@{}}Pruning \\ {[}10, 20, 30{]}\%\end{tabular}} \\ \cline{2-9} 
\multicolumn{1}{|l|}{}                                & \multicolumn{2}{c|}{\textbf{MNIST (Gaussian, 0.01)}}                                  & \multicolumn{2}{c|}{\textbf{CIFAR (Gaussian, 0.01)}}                                  & \multicolumn{2}{c|}{\textbf{MNIST (Gaussian, 0.01)}}                                                                          & \multicolumn{2}{c|}{\textbf{MNIST (S\&P noise, 10\%)}}                                                                                                                                        \\ \hline
\multicolumn{1}{|l|}{\textbf{SSIM}}                   & \multicolumn{2}{c|}{{[}0.80, 0.78, 0.75, 0.57, 0.05{]}}                               & \multicolumn{2}{c|}{{[}0.76, 0.75, 0.73, 0.36, -0.005{]}}                             & \multicolumn{2}{c|}{{[}0.76, 0.71, 0.68, 0.62{]}}                                                                             & \multicolumn{1}{c|}{{[}0.69,0.65,0.63{]}}                                                           & {[}0.69,0.65,0.62{]}                                                           \\ \hline
\multicolumn{1}{|l|}{\textbf{MSE (x10)}}              & \multicolumn{2}{c|}{{[}0.03, 0.04, 0.06, 0.06, 60.0{]}}                               & \multicolumn{2}{c|}{{[}0.05, 0.05, 0.06, 0.63, 72.1{]}}                               & \multicolumn{2}{c|}{{[}0.06, 0.13, 0.25, 0.62{]}}                                                                             & \multicolumn{1}{c|}{{[}0.12,0.18,0.24{]}}                                                           & {[}0.11,0.18,0.25{]}                                                           \\ \hline
\multicolumn{1}{|l|}{\textbf{PSNR (dB)}}              & \multicolumn{2}{c|}{{[}24.2, 23.7, 21.8, 12.5,-7.5{]}}                                & \multicolumn{2}{c|}{{[}21.8, 21.7, 20.9, 11.4, -9.1{]}}                               & \multicolumn{2}{c|}{{[}22.0, 18.9, 16.1, 12.3{]}}                                                                             & \multicolumn{1}{c|}{{[}19.2,17.5,16.3{]}}                                                           & {[}18.7,17.3,16.1{]}                                                           \\ \hline
\end{tabular}
\label{newtable}
\vspace{-0.5cm}
\end{table*}

\vspace{-0.2cm}
\subsection{Denoising  network architecture}
\vspace{-0.15cm}
The denoising network for near-sensor processing and retraining for different types of noises can be implemented in two ways: (1) single-layered dense network and (2) multi-layered convolution/deconvolution network (CNN). Both approaches, as shown in Fig. \ref{f01}, have advantaged and drawbacks. 
In both approaches, positive and negative weights of the network are implemented with two memristors and opamp based readouts. Also, we assume crossbars to be separated to 256$\times$64 crossbar tiles \cite{wang2021taichi}.

\subsubsection{Dense Network for denoising}
Fig. \ref{f01} (a) illustrates the training of a single-layer fully-connected denoising network. Comparing to backpropagation, retraining of such a network is less complicated. It is based on root mean square error (RMSE) calculation between the ideal and real outputs and updating RRAM weights according to this error. It involves multiplication of error with the corresponding inputs. Such a network is smaller and consumes less energy due to simplified training. But as in all shallow networks, the number of devices to train is higher than in a CNN when aiming for the same inference performance (Table \ref{t_hardware}).







\subsubsection{Training of CNN denoising network}
Fig. \ref{f01} (b) illustrates more complex denoising network involving convolution and deconvolution parts. The noisy image is applied to a convolution layer that extracts useful features, and the pooling layer reduces the dimensions of the images. Then, the 3D convolution collects all the feature maps from the previous layer into a single image. Another convolution layer and up-sampling reconstruct the image back, reducing the noise in the images. 
As convolution filters are small, the number of trained devices is around 70 times less than in a fully connected denoising network, if implemented with a sequential readout (Table \ref{t_hardware}). This can reduce the fabrication cost of a crossbar part. However, the network is spread-out, and complex mixed-signal backpropagation circuits are required to retrain the network, which is computationally expensive for edge devices. Also, the outputs of intermediate layers should be stored for error calculation. In Fig. \ref{f01} (b), it is implemented using ADCs and SRAM contributing to area and power overhead. As same convolution kernels are used to process different parts of an image, the adder/accumulator is required to sum up the corresponding errors when propagating back before the update.

\vspace{-0.15cm}
\section{Simulation results and comparison}

\vspace{-0.15cm}

\subsubsection{Denoising performance}
The proposed system has been tested for MNIST \cite{lecun1998gradient} and CIFAR-10 \cite{krizhevsky2009learning}  datasets (Fig.\ref{newfigure} (a-b))  for eight different input noises \cite{boncelet2009image}: Gaussian noise with $\sigma$ =$0.01,0.1,0.5$, Salt and Pepper (S\&P) noise of 10\%, 25\% and 50\%, Poisson, and Speckle noise shown. In Table \ref{newtable}, Structural similarity index (SSIM) \cite{wang2004image}, mean square error (MSE) and Peak Signal-to-Noise Ratio (PSNR) have been measured for all the images calculating the average score for the whole testing set. For MNIST dataset, the dense network results are slightly better than CNN, as convolution and pooling compresses some structural information. Denoised images have higher SSIM improved up to 2.28 and 3.61 times for MNIST and CIFAR images. The gap between SSIM of noisy and denoised images becomes more significant for the case where structural information is nearly destroyed by noise (e.g. 
Gaussian noise with $\sigma$=0.5 or S\&P noise of 50\%). Mostly, MSE of the denoised images is 10 times lower than of the noisy ones, which can be up to 16.3 and 21.7 times lower for MNIST and CIFAR datasets for high Gaussian noise. On average, PSNR is increased by 5-10dB for images with medium noise and up to 20dB for images with high noise.
Maximum PSNR improvement is 5.1 and 7.7 times for MNIST and CIFAR images.

\vspace{-0.05cm}
\subsubsection{Device non-idealities, dropout and pruning}
Fig. \ref{newfigure} (c-d) and Table \ref{newtable}
illustrates how the denoised image's quality is affected by memristor non-idealities \cite{krestinskaya2019memristive}, dropout and pruning. For limited number of stable resistive states $L<16$, denoised image quality is significantly reduced and is worse than the noisy images.
The desirable number of states is $L$=$64$-$128$ for MNIST and $L$=$128$-$256$ for CIFAR.
The conductance variation with $\sigma$=$0.025$ 
reduces SSIM and PSNR by 18\% and 40\%, respectively. 
Fig. \ref{newfigure} (e) demonstrates how dropout (randomly dropping out input neurons\cite{krestinskaya2020analogue}) and pruning (randomly disconnecting memristor devices introducing sparsity\cite{guo2020unsupervised}) can be used to reduce the number of active devices in the fully connected denoising layer.
The dropout and pruning of 20\% devices cause around 2dB reduction of PSNR (Table \ref{newtable}).




\begin{table}[]
\centering
\caption{\small Comparison with conventional denoising methods \cite{goyal2020image}.}
\scalebox{.7}{
\begin{tabular}{|m{2cm}|m{0.55cm}|m{0.55cm}|m{0.55cm}|m{0.55cm}|m{0.55cm}|m{0.55cm}|m{0.55cm}|m{0.55cm}|}
\hline
\multirow{2}{*}{\textbf{Method}}                                              & \multicolumn{3}{c|}{\textbf{Gaussian}}                                                                                                                                                                                      & \multicolumn{3}{c|}{\textbf{Salt and Pepper}}                                                                                                                                                                                & \multirow{2}{*}{\textbf{\begin{tabular}[c]{@{}c@{}}Pois-\\ son\end{tabular}}} & \multirow{2}{*}{\textbf{\begin{tabular}[c]{@{}c@{}}Spe-\\ ckle\end{tabular}}} \\ \cline{2-7}
                                                                     & \textbf{0.01}                                                           & \textbf{0.1}                                                            & \textbf{0.5}                                                            & \textbf{0.1}                                                            & \textbf{0.25}                                                            & \textbf{0.5}                                                            &                                                                               &                                                                              \\ \hline
\begin{tabular}[c]{@{}l@{}}\textbf{No filter} \textbf{ - Noisy }\\ SSIM:\\ MSE:\\ PSNR:\end{tabular} & \begin{tabular}[c]{@{}c@{}}\\0.719\\ 0.009\\ 20.0dB\end{tabular}          & \begin{tabular}[c]{@{}c@{}}\\0.489\\ 0.099\\ 10.0dB\end{tabular}          &  \begin{tabular}[c]{@{}c@{}}\\0.232\\ 0.499\\ 3.0dB\end{tabular}           & \begin{tabular}[c]{@{}c@{}}\\0.607\\ 0.004\\ -\end{tabular}               & \begin{tabular}[c]{@{}c@{}}\\0.449\\ 0.101\\ -\end{tabular}                & \begin{tabular}[c]{@{}c@{}}\\0.317\\ 0.171\\ -\end{tabular}               & \begin{tabular}[c]{@{}c@{}}\\0.832\\ 0.052\\ 13.1dB\end{tabular}                & \begin{tabular}[c]{@{}c@{}}\\0.684\\ 0.114\\ 9.7dB\end{tabular}                \\ \hline
\begin{tabular}[c]{@{}l@{}}\textbf{This work}\\ SSIM:\\ MSE:\\ PSNR:\end{tabular}        & \cellcolor{green!10} \textbf{\begin{tabular}[c]{@{}c@{}}\\0.803\\ 0.003\\ 24.3dB\end{tabular}} & \cellcolor{green!10} \textbf{\begin{tabular}[c]{@{}c@{}}\\0.673\\ 0.014\\ 17.5dB\end{tabular}} & \cellcolor{yellow!15} \textbf{\begin{tabular}[c]{@{}c@{}}\\0.529\\ 0.030\\ 15.3dB\end{tabular}}  & \cellcolor{green!10} \textbf{\begin{tabular}[c]{@{}c@{}}\\0.708\\ 0.009\\ 20.2dB\end{tabular}}  & \cellcolor{yellow!15} \textbf{\begin{tabular}[c]{@{}c@{}}\\0.640\\ 0.016\\ 17.9dB\end{tabular}}  & \cellcolor{yellow!15} \textbf{\begin{tabular}[c]{@{}c@{}}\\0.571\\ 0.025\\ 16.1dB\end{tabular}} &\cellcolor{yellow!15} \textbf{\begin{tabular}[c]{@{}c@{}}\\0.709\\ 0.013\\ 18.9dB\end{tabular} }               & \cellcolor{yellow!15} \textbf{\begin{tabular}[c]{@{}c@{}}\\0.749\\ 0.018\\ 17.2dB\end{tabular}}      \\ \hline
\begin{tabular}[c]{@{}l@{}}\textbf{Gaussian filter}\\ SSIM:\\ MSE:\\ PSNR:\end{tabular}       & \begin{tabular}[c]{@{}c@{}}\\0.748\\ 0.006\\ 22.3dB\end{tabular}          & \begin{tabular}[c]{@{}c@{}}\\0.58\\ 0.022\\ 16.6dB\end{tabular}           & \cellcolor{green!10} \textbf{\begin{tabular}[c]{@{}c@{}}\\0.398\\ 0.040\\ 13.3dB\end{tabular}}  & \begin{tabular}[c]{@{}c@{}}\\0.621\\ 0.018\\ 17.6dB\end{tabular}          &  \begin{tabular}[c]{@{}c@{}}\\0.521\\ 0.034\\ 14.6dB\end{tabular}  & \begin{tabular}[c]{@{}c@{}}\\0.411\\ 0.062\\ 12.0dB\end{tabular}          & \cellcolor{green!10} \textbf{\begin{tabular}[c]{@{}c@{}}\\0.844\\ 0.014\\ 18.5dB\end{tabular}}     & \cellcolor{green!10} \textbf{\begin{tabular}[c]{@{}c@{}}\\0.748\\ 0.022\\ 16.7dB\end{tabular}}      \\ \hline
\begin{tabular}[c]{@{}l@{}}\textbf{Median filter}\\ SSIM:\\ MSE:\\ PSNR:\end{tabular}             & \begin{tabular}[c]{@{}c@{}}\\0.742\\ 0.007\\ 21.5dB\end{tabular}          & \begin{tabular}[c]{@{}c@{}}\\0.571\\ 0.027\\ 15.5dB\end{tabular}          & \begin{tabular}[c]{@{}c@{}}\\0.331\\ 0.067\\ 11.8dB\end{tabular}          & \cellcolor{yellow!15} \textbf{\begin{tabular}[c]{@{}c@{}}\\0.914\\ 0.008\\ 21.1dB\end{tabular}} & \cellcolor{yellow!15} \textbf{\begin{tabular}[c]{@{}c@{}}\\0.842\\ 0.017\\ 17.9dB\end{tabular}} & \cellcolor{green!10} \textbf{\begin{tabular}[c]{@{}c@{}}\\0.317\\ 0.040\\ 14.1dB\end{tabular}}          & \begin{tabular}[c]{@{}c@{}}\\0.749\\ 0.025\\ 16.0dB\end{tabular}      & \begin{tabular}[c]{@{}c@{}}\\0.699\\ 0.030\\ 14.8dB\end{tabular}               \\ \hline
\begin{tabular}[c]{@{}l@{}}\textbf{Bilateral filter}\\ SSIM:\\ MSE:\\ PSNR:\end{tabular}     & \begin{tabular}[c]{@{}c@{}}\\0.678\\ 0.008\\ 20.8dB\end{tabular}          & \begin{tabular}[c]{@{}c@{}}\\0.503\\ 0.031\\ 15.0dB\end{tabular}          & \begin{tabular}[c]{@{}c@{}}\\0.288\\ 0.116\\ 9.3dB\end{tabular}           & \begin{tabular}[c]{@{}c@{}}\\0.638\\ 0.027\\ 15.7dB\end{tabular}          & \begin{tabular}[c]{@{}c@{}}\\0.469\\ 0.057\\ 12.4dB\end{tabular}           & \begin{tabular}[c]{@{}c@{}}\\0.322\\ 0.088\\ 10.5dB\end{tabular}          & \begin{tabular}[c]{@{}c@{}}\\0.714\\ 0.045\\ 13.6dB\end{tabular}                & \begin{tabular}[c]{@{}c@{}}\\0.648\\ 0.067\\ 11.9dB\end{tabular}               \\ \hline

\begin{tabular}[c]{@{}l@{}}\textbf{NL means filter}\\ SSIM:\\ MSE:\\ PSNR:\end{tabular} & \cellcolor{yellow!15}\textbf{\begin{tabular}[c]{@{}c@{}}\\0.913\\ 0.002\\ 27.2dB\end{tabular}} & \cellcolor{yellow!15} \textbf{\begin{tabular}[c]{@{}c@{}}\\0.696\\ 0.015\\ 18.3dB\end{tabular}} &  \begin{tabular}[c]{@{}c@{}}\\0.351\\ 0.057\\ 12.5dB\end{tabular}          & \cellcolor{green!10} \textbf{\begin{tabular}[c]{@{}c@{}}\\0.686\\ 0.010\\ 19.9dB\end{tabular}} & \cellcolor{green!10} \textbf{\begin{tabular}[c]{@{}c@{}}\\0.456\\ 0.032\\ 15.0dB\end{tabular}}           & \begin{tabular}[c]{@{}c@{}}\\0.345\\ 0.057\\ 12.4dB\end{tabular}          & \begin{tabular}[c]{@{}c@{}}\\0.726\\ 0.044\\ 13.7dB\end{tabular}               & \begin{tabular}[c]{@{}c@{}}\\0.679\\ 0.111\\ 9.8dB\end{tabular}                \\ \hline
\begin{tabular}[c]{@{}l@{}}\textbf{Wavelet filter}\\ SSIM:\\ MSE:\\ PSNR:\end{tabular}            & \begin{tabular}[c]{@{}c@{}}\\0.753\\ 0.005\\ 22.4dB\end{tabular}          & \begin{tabular}[c]{@{}c@{}}\\0.559\\ 0.038\\ 14.2dB\end{tabular}          & \begin{tabular}[c]{@{}c@{}}\\0.346\\ 0.141\\ 8.5dB\end{tabular}           & \begin{tabular}[c]{@{}c@{}}\\0.616\\ 0.033\\ 14.7dB\end{tabular}          & \begin{tabular}[c]{@{}c@{}}\\0.482\\ 0.054\\ 13.2dB\end{tabular}           & \begin{tabular}[c]{@{}c@{}}\\0.368\\ 0.078\\ 11.0dB\end{tabular} & \begin{tabular}[c]{@{}c@{}}\\0.765\\ 0.029\\ 15.5dB\end{tabular}      & \begin{tabular}[c]{@{}c@{}}\\0.712\\ 0.04\\ 13.7dB\end{tabular}                \\ \hline

\begin{tabular}[c]{@{}l@{}}\textbf{Total variation }\\ SSIM:\\ MSE:\\ PSNR:\end{tabular}   & \cellcolor{green!10} \textbf{\begin{tabular}[c]{@{}c@{}}\\0.849\\ 0.002\\ 26.2dB\end{tabular}} & \cellcolor{green!10} \textbf{\begin{tabular}[c]{@{}c@{}}\\0.624\\ 0.015\\ 18.1dB\end{tabular}} & \cellcolor{green!10} \textbf{\begin{tabular}[c]{@{}c@{}}\\0.394\\ 0.043\\ 13.6dB\end{tabular}} & \begin{tabular}[c]{@{}c@{}}\\0.614\\ 0.025\\ 16.0dB\end{tabular}          & \begin{tabular}[c]{@{}c@{}}\\0.482\\ 0.034\\ 14.6dB\end{tabular}           & \begin{tabular}[c]{@{}c@{}}\\0.360\\ 0.060\\ 12.1dB\end{tabular}          & \cellcolor{green!10} \textbf{\begin{tabular}[c]{@{}c@{}}\\0.743\\ 0.021\\ 16.9dB\end{tabular} }              &  \begin{tabular}[c]{@{}c@{}}\\0.599\\ 0.032\\ 15.1dB\end{tabular}               \\ \hline
\end{tabular}}
\label{t_comp}
\end{table}

\vspace{-0.05cm}
\subsubsection{Comparison to the conventional methods}
Table \ref{t_comp} shows the comparison of dense network performance with conventional denoising methods \cite{goyal2020image} for MNIST dataset, where top 1 result is highlighted in yellow and top 3 are highlighted in green. For conventional methods, the highest scores from several iterations are considered. The proposed method is more generalized showing stable performance and is among top 3 best results for all types of noises. 

\begin{table*}[]
\scriptsize
\caption{\small  Hardware performance estimation for inference and training.}
\centering
\begin{tabular}{|llccccccccccc|}
\hline
\multicolumn{2}{|l|}{\multirow{3}{*}{\textbf{Network}}}                  & \multicolumn{11}{c|}{\cellcolor{green!10}\textbf{Inference}}                                                                                                                                                                                                                                                                                                                                                                                                                                                                                                                                                                                                   \\ \cline{3-13} 
\multicolumn{2}{|l|}{}                                                   & \multicolumn{1}{c|}{\multirow{2}{*}{\textbf{\begin{tabular}[c]{@{}c@{}}Number of \\ devices\end{tabular}}}} & \multicolumn{1}{c|}{\multirow{2}{*}{\textbf{tiles}}} & \multicolumn{3}{c|}{\textbf{Crossbar}}                                                                                               & \multicolumn{3}{c|}{\textbf{CMOS}}                                                                                                            & \multicolumn{2}{c|}{\textbf{Total}}                                                          & \multirow{2}{*}{\textbf{\begin{tabular}[c]{@{}c@{}}Time per \\ image\end{tabular}}} \\ \cline{5-12}
\multicolumn{2}{|l|}{}                                                   & \multicolumn{1}{c|}{}                                                                                   & \multicolumn{1}{c|}{}                                & \multicolumn{1}{c|}{\textbf{Power}} & \multicolumn{1}{c|}{\textbf{Energy}}          & \multicolumn{1}{c|}{\textbf{Area}}             & \multicolumn{1}{c|}{\textbf{Power}}          & \multicolumn{1}{c|}{\textbf{Energy}}          & \multicolumn{1}{c|}{\textbf{Area}}             & \multicolumn{1}{c|}{\textbf{Energy}}         & \multicolumn{1}{c|}{\textbf{Area}}            &                                                                                     \\ \hline
\multicolumn{1}{|l|}{\multirow{2}{*}{Dense}} & \multicolumn{1}{l|}{seq.} & \multicolumn{1}{c|}{\multirow{2}{*}{614625x2}}                                                          & \multicolumn{1}{c|}{\multirow{2}{*}{52x2}}           & \multicolumn{1}{c|}{0.222$mW$}        & \multicolumn{1}{c|}{\multirow{2}{*}{0.577$nJ$}} & \multicolumn{1}{c|}{\multirow{2}{*}{0.291$mm^2$}} & \multicolumn{1}{c|}{6.428$mW$}                 & \multicolumn{1}{c|}{\multirow{2}{*}{20.57$nJ$}} & \multicolumn{1}{c|}{0.0006$mm^2$}                 & \multicolumn{1}{l|}{\multirow{2}{*}{21.1$nJ$}} & \multicolumn{1}{l|}{0.29$mm^2$}                  & 3.2$\mu s$                                                                              \\ \cline{2-2} \cline{5-5} \cline{8-8} \cline{10-10} \cline{12-13} 
\multicolumn{1}{|l|}{}                       & \multicolumn{1}{l|}{par.} & \multicolumn{1}{c|}{}                                                                                   & \multicolumn{1}{c|}{}                                & \multicolumn{1}{c|}{11.54$mW$}        & \multicolumn{1}{c|}{}                         & \multicolumn{1}{c|}{}                          & \multicolumn{1}{c|}{411.4$mW$}                 & \multicolumn{1}{c|}{}                         & \multicolumn{1}{c|}{0.0404$mm^2$}                 & \multicolumn{1}{l|}{}                        & \multicolumn{1}{l|}{0.33$mm^2$}                  &   50ns                                                                               \\ \hline
\multicolumn{1}{|l|}{\multirow{2}{*}{CNN}}   & \multicolumn{1}{l|}{seq.} & \multicolumn{1}{c|}{8834x2}                                                                             & \multicolumn{1}{c|}{6x2}                             & \multicolumn{1}{c|}{1.332$mW$}        & \multicolumn{1}{c|}{70.86$nJ$}                  & \multicolumn{1}{c|}{0.033$mm^2$}                  & \multicolumn{1}{c|}{202.5$mW$}                 & \multicolumn{1}{c|}{65.32$\mu J$}                  & \multicolumn{1}{c|}{0.0198$mm^2$}                 & \multicolumn{1}{l|}{65.4$\mu J$}                  & \multicolumn{1}{l|}{0.05$mm^2$}                  & 319$\mu s$                                                                              \\ \cline{2-13} 
\multicolumn{1}{|l|}{}                       & \multicolumn{1}{l|}{par.} & \multicolumn{1}{c|}{3840032x2}                                                                          & \multicolumn{1}{c|}{1116x2}                          & \multicolumn{1}{c|}{247.7$mW$}        & \multicolumn{1}{c|}{1.486$n J$}                  & \multicolumn{1}{c|}{6.249$mm^2$}                  & \multicolumn{1}{c|}{37.68$W$}                  & \multicolumn{1}{c|}{11.34$\mu J$}                  & \multicolumn{1}{c|}{3.7051$mm^2$}                 & \multicolumn{1}{l|}{11.3$\mu J$}                  & \multicolumn{1}{l|}{9.93$mm^2$}                  & 0.3$\mu s$                                                                               \\ \hline
\multicolumn{2}{|l|}{\multirow{3}{*}{\textbf{Network}}}                  & \multicolumn{11}{c|}{\cellcolor{green!10}\textbf{Training}}                                                                                                                                                                                                                                                                                                                                                                                                                                                                                                                                                                                                    \\ \cline{3-13} 
\multicolumn{2}{|l|}{}                                                   & \multicolumn{2}{c|}{\multirow{2}{*}{\textbf{\begin{tabular}[c]{@{}c@{}}Intermediate\\ output\end{tabular}}}}                                                   & \multicolumn{3}{c|}{\textbf{SRAM + ADC}}                                                                                             & \multicolumn{3}{c|}{\textbf{CMOS}}                                                                                                            & \multicolumn{2}{c|}{\textbf{Total}}                                                          & \multirow{2}{*}{\textbf{\begin{tabular}[c]{@{}c@{}}Time per\\  image\end{tabular}}} \\ \cline{5-12}
\multicolumn{2}{|l|}{}                                                   & \multicolumn{2}{c|}{}                                                                                                                                          & \multicolumn{1}{c|}{\textbf{Power}} & \multicolumn{1}{c|}{\textbf{Energy}}          & \multicolumn{1}{c|}{\textbf{Area}}             & \multicolumn{1}{c|}{\textbf{Power}}          & \multicolumn{1}{c|}{\textbf{Energy}}          & \multicolumn{1}{c|}{\textbf{Area}}             & \multicolumn{1}{c|}{\textbf{Energy}}         & \multicolumn{1}{c|}{\textbf{Area}}            &                                                                                     \\ \hline
\multicolumn{2}{|l|}{\multirow{2}{*}{Dense}}                             & \multicolumn{2}{c|}{\multirow{2}{*}{-}}                                                                                                                        & \multicolumn{3}{c|}{\multirow{2}{*}{-}}                                                                                              & \multicolumn{1}{c|}{\multirow{2}{*}{32.8$mW$}} & \multicolumn{1}{c|}{\multirow{2}{*}{214.9$\mu J$}} & \multicolumn{1}{c|}{\multirow{2}{*}{0.481$mm^2$}} & \multicolumn{1}{c|}{\multirow{2}{*}{236$\mu J$}}  & \multicolumn{1}{c|}{\multirow{2}{*}{0.48$mm^2$}} & \multirow{2}{*}{72$ms$}                                                               \\
\multicolumn{2}{|l|}{}                                                   & \multicolumn{2}{c|}{}                                                                                                                                          & \multicolumn{3}{c|}{}                                                                                                                & \multicolumn{1}{c|}{}                        & \multicolumn{1}{c|}{}                         & \multicolumn{1}{c|}{}                          & \multicolumn{1}{c|}{}                        & \multicolumn{1}{c|}{}                         &                                                                                     \\ \hline
\multicolumn{1}{|l|}{\multirow{2}{*}{CNN}}   & \multicolumn{1}{l|}{seq.} & \multicolumn{2}{c|}{\multirow{2}{*}{84868}}                                                                                                                    & \multicolumn{1}{c|}{0.24$mW$}         & \multicolumn{1}{c|}{4.90$\mu J$}                   & \multicolumn{1}{c|}{0.018$mm^2$}                  & \multicolumn{1}{c|}{82.1$mW$}                  & \multicolumn{1}{c|}{261.9$\mu J$}                  & \multicolumn{1}{c|}{0.486$mm^2$}                  & \multicolumn{1}{c|}{331$\mu J$}                   & \multicolumn{1}{c|}{0.50$mm^2$}                  & 41.9$s$                                                                               \\ \cline{2-2} \cline{5-13} 
\multicolumn{1}{|l|}{}                       & \multicolumn{1}{l|}{par.} & \multicolumn{2}{c|}{}                                                                                                                                          & \multicolumn{1}{c|}{44.6$mW$}         & \multicolumn{1}{c|}{0.85$\mu J$}                   & \multicolumn{1}{c|}{3.348$mm^2$}                  & \multicolumn{1}{c|}{150.0$W$}                    & \multicolumn{1}{c|}{451.2$\mu J$}                  & \multicolumn{1}{c|}{675.1$mm^2$}                  & \multicolumn{1}{c|}{463$\mu J$}                   & \multicolumn{1}{c|}{678$mm^2$}                   & 74$ms$                                                                                \\ \hline
\end{tabular}
\label{t_hardware}
\vspace{-0.5cm}
\end{table*}

\vspace{-0.05cm}
\subsubsection{Hardware performance estimation} Table \ref{t_hardware} shows on-chip area, power consumption, processing time and energy per image denoising for inference and training circuits in 65nm CMOS technology. We take into account measurements data from 256$\times$64 crossbar tiles with power consumption of $111\mu W$ and on-chip area of $2800 \mu m^2$ per tile \cite{wang2021taichi}. A single read and write operation is estimated as $50ns$ and $80ns$ (12.5-20MHz) with 50 update pulses on average \cite{liu2020neural,cai2019fully}.
The sequential and parallel operations are considered for both networks.
In sequential inference in dense network, the outputs in each tile share a single readout circuit, while in parallel processing each output is connected to a separate readout. In dense network, we consider opamp based summation circuits to sum up the outputs from several tiles. In CNN design, we map CNN kernels to the crossbars by unfolding and duplicating weights \cite{wang2019deep,lammie2019variation}. In parallel processing in CNN, the convolution kernels are duplicated, which require 400 times more RRAM devices. In CNN, a single readout and ADC per tile is assumed.
The opamp design adapted from \cite{saxena2010indirect} consumes $262.4\mu W$ of power and $25.8\mu m^2$ of area. We use an $1.731\mu m^2$/$0.48 \mu W$ analog current multiplier \cite{danesh2019ultra}, 
adding opamp based voltage-to-current converters, which makes the area and power of a multiplier $53.33\mu m^2$ and $525.28\mu W$. For storage of intermediate outputs, we assume 8-bit $3000\mu m^2$/$40\mu W$ ADC \cite{wang2021taichi} and 6T SRAM cells of $0.525\mu m^2$ \cite{khwa201865nm}. The pooling is CNN is performed by an opamp-based max-pooling circuit \cite{yildirim2021analog}.

In the training circuits, we consider a single difference amplifier, multiplier, ADC per output crossbar tile in both networks. The training time is estimated considering forward and backward propagation, CMOS processing and update time. 
All crossbar tiles are updated in parallel sequentially inside the tile. We do not consider digital control circuits, DACs and adders in CNN design, assuming them to be a part of the control circuits overhead calculation.
The energy consumption of the proposed dense denoising network  
is the same as of a digital pre-processor for median filtering and 22 times lower than for NL Means filtering for MNIST images \cite{wanimage}. Also, the dense network with parallel processing has 47 times smaller area and 22 times lower power consumption
than state-of-the-art  65nm 8-bit image denoising accelerator \cite{mahmoud2017ideal}.



\squeezeupupupi
\begin{figure}[!h]
    \centering        
    \includegraphics[width=80mm]{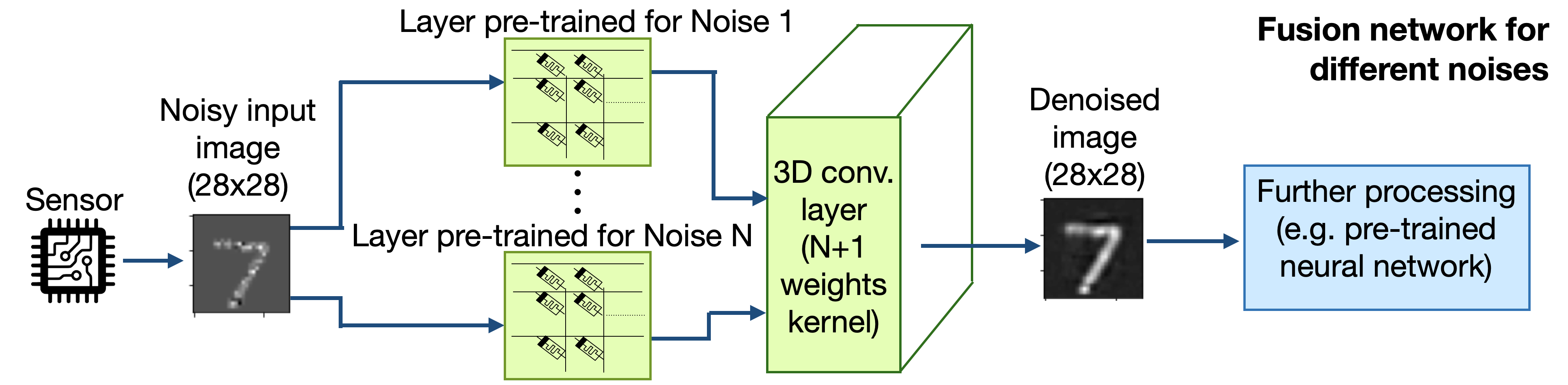}
    \caption{\footnotesize Network fusion approach for denoising images under unknown variable noise conditions.}
    \label{f30}
\end{figure}

\squeezeupupupi

\begin{figure}[!h]
    \centering        
   \includegraphics[width=90mm]{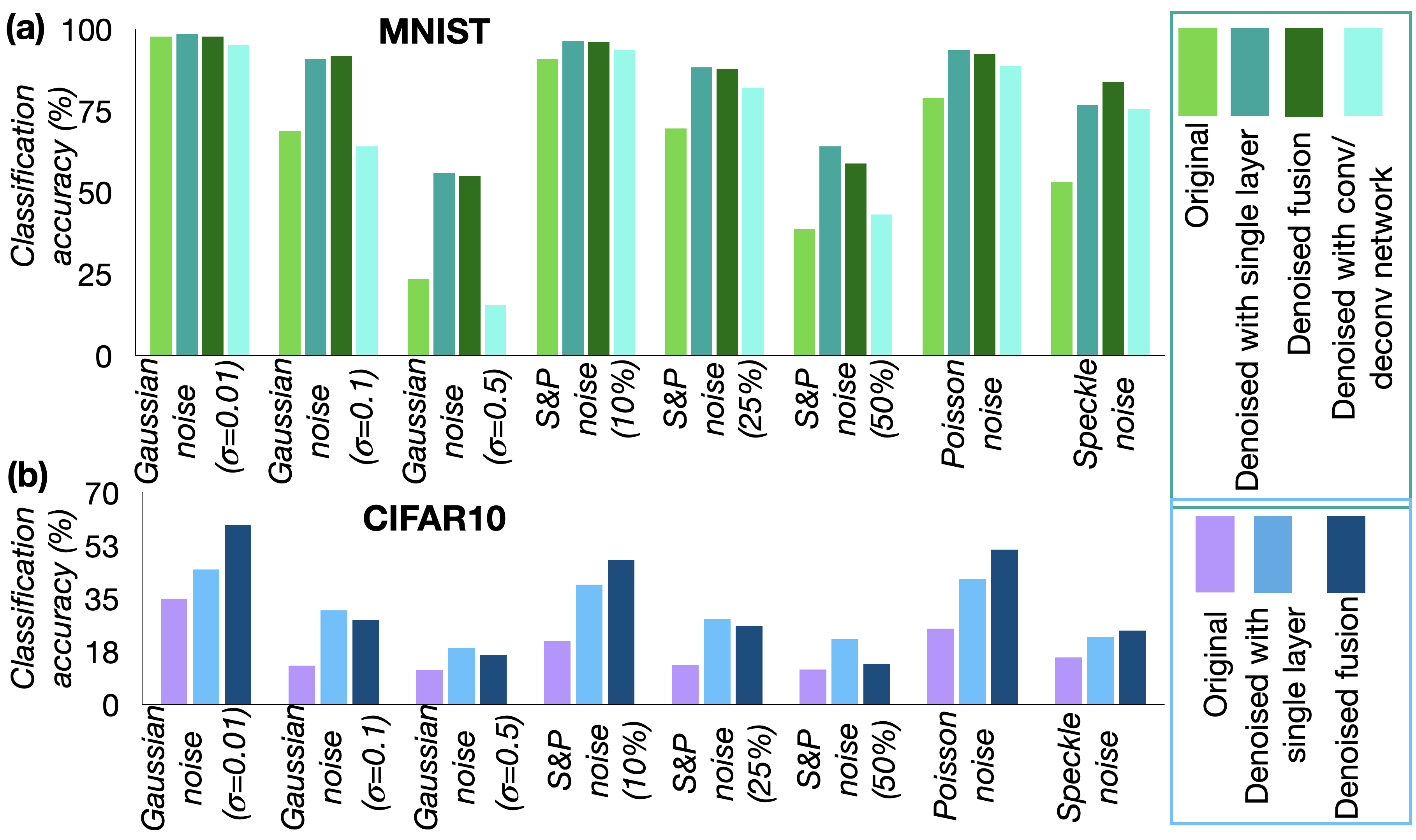}
    \caption{\footnotesize Classification accuracy improvement with denoising for different types of noises for (a) MNIST and (b) CIFAR10.}
    \label{f3}
\end{figure}

\subsubsection{Application to image classification}

After reading the sensor's output and enhancing image quality, the denoised image can be used in pre-trained networks for image classification. If the noise is unknown, there is no unique method to filter out all types of noises. To address this issue, we introduce the fusion network shown in Fig. \ref{f30}. The same noisy image from the sensor is applied to several denoising networks trained for different noises. The outputs of all denoising networks are fused using 3D convolution into a single denoised image. The 3D convolution layer is trained separately using the same RMSE approach. We tested the fusion approach for eight pre-trained denoising dense networks and a 3D convolution layer trained with a set of 50\% of noiseless images and 50\% of images with different noises.
The simulation results are illustrated in Fig. \ref{f3}.
For MNIST database, the classification accuracy is increased by 30\% for high Gaussian noise when introducing denoising network and by 20\% for high S\&P noise and Speckle noise. The performance of the fusion network is comparable to the performance of separate single-layer denoising networks. This shows that fusion is a useful technique when the noise type is unknown, or the conditions change in real-time. 

\vspace{-0.1cm}

\section{Conclusion}
\vspace{-0.1cm}

We proposed adaptive denoising memristive networks that can
integrate into analog image sensors and trained on-chip
adapting to new noisy conditions. This approach allows retraining only a small denoising network without retraining
the entire system for new noisy conditions. For adapting to
different types of noises, the fusion approach is introduced. The method is verified for image classification. The application scope of the proposed system can be extended to different problems that require real-time image denoising as a pre-processing step.


\balance
\bibliographystyle{IEEEtran}
\bibliography{ref.bib}

\begin{thebibliography}{10}
\providecommand{\url}[1]{#1}
\csname url@samestyle\endcsname
\providecommand{\newblock}{\relax}
\providecommand{\bibinfo}[2]{#2}
\providecommand{\BIBentrySTDinterwordspacing}{\spaceskip=0pt\relax}
\providecommand{\BIBentryALTinterwordstretchfactor}{4}
\providecommand{\BIBentryALTinterwordspacing}{\spaceskip=\fontdimen2\font plus
\BIBentryALTinterwordstretchfactor\fontdimen3\font minus
  \fontdimen4\font\relax}
\providecommand{\BIBforeignlanguage}[2]{{%
\expandafter\ifx\csname l@#1\endcsname\relax
\typeout{** WARNING: IEEEtran.bst: No hyphenation pattern has been}%
\typeout{** loaded for the language `#1'. Using the pattern for}%
\typeout{** the default language instead.}%
\else
\language=\csname l@#1\endcsname
\fi
#2}}
\providecommand{\BIBdecl}{\relax}
\BIBdecl

\bibitem{el2005cmos}
A.~El~Gamal and H.~Eltoukhy, ``Cmos image sensors,'' \emph{IEEE Circuits and
  Devices Magazine}, vol.~21, no.~3, pp. 6--20, 2005.

\bibitem{gow2007comprehensive}
R.~D. Gow, D.~Renshaw, K.~Findlater, L.~Grant, S.~J. McLeod, J.~Hart, and R.~L.
  Nicol, ``A comprehensive tool for modeling cmos image-sensor-noise
  performance,'' \emph{IEEE Transactions on Electron Devices}, vol.~54, no.~6,
  pp. 1321--1329, 2007.

\bibitem{oike2012cmos}
Y.~Oike and A.~El~Gamal, ``Cmos image sensor with per-column $\sigma$ $\delta$
  adc and programmable compressed sensing,'' \emph{IEEE Journal of Solid-State
  Circuits}, vol.~48, no.~1, pp. 318--328, 2012.

\bibitem{massari2005cmos}
N.~Massari, M.~Gottardi, L.~Gonzo, D.~Stoppa, and A.~Simoni, ``A cmos image
  sensor with programmable pixel-level analog processing,'' \emph{IEEE
  Transactions on Neural Networks}, vol.~16, no.~6, pp. 1673--1684, 2005.

\bibitem{chen2019processing}
Z.~Chen, H.~Zhu, E.~Ren, Z.~Liu, K.~Jia, L.~Luo, X.~Zhang, Q.~Wei, F.~Qiao,
  X.~Liu \emph{et~al.}, ``Processing near sensor architecture in mixed-signal
  domain with cmos image sensor of convolutional-kernel-readout method,''
  \emph{IEEE Transactions on Circuits and Systems I: Regular Papers}, vol.~67,
  no.~2, pp. 389--400, 2019.

\bibitem{nazare2017deep}
T.~S. Nazar{\'e}, G.~B.~P. da~Costa, W.~A. Contato, and M.~Ponti, ``Deep
  convolutional neural networks and noisy images,'' in \emph{Iberoamerican
  Congress on Pattern Recognition}.\hskip 1em plus 0.5em minus 0.4em\relax
  Springer, 2017, pp. 416--424.

\bibitem{diamond2017dirty}
S.~Diamond, V.~Sitzmann, S.~Boyd, G.~Wetzstein, and F.~Heide, ``Dirty pixels:
  Optimizing image classification architectures for raw sensor data,''
  \emph{arXiv preprint arXiv:1701.06487}, 2017.

\bibitem{kawahito2008cmos}
S.~Kawahito, J.-H. Park, K.~Isobe, S.~Shafie, T.~Iida, and T.~Mizota, ``A cmos
  image sensor integrating column-parallel cyclic adcs with on-chip digital
  error correction circuits,'' in \emph{2008 IEEE International Solid-State
  Circuits Conference-Digest of Technical Papers}.\hskip 1em plus 0.5em minus
  0.4em\relax IEEE, 2008, pp. 56--595.

\bibitem{xu2020macsen}
H.~Xu, Z.~Li, N.~Lin, Q.~Wei, F.~Qiao, X.~Yin, and H.~Yang, ``Macsen: A
  processing-in-sensor architecture integrating mac operations into image
  sensor for ultra-low-power bnn-based intelligent visual perception,''
  \emph{IEEE Transactions on Circuits and Systems II: Express Briefs}, 2020.

\bibitem{krestinskaya2020memristive}
O.~Krestinskaya, B.~Choubey, and A.~James, ``Memristive gan in analog,''
  \emph{Scientific Reports}, vol.~10, no.~1, pp. 1--14, 2020.

\bibitem{krestinskaya2018analog}
O.~Krestinskaya, K.~N. Salama, and A.~P. James, ``Analog backpropagation
  learning circuits for memristive crossbar neural networks,'' in \emph{2018
  IEEE International Symposium on Circuits and Systems (ISCAS)}.\hskip 1em plus
  0.5em minus 0.4em\relax IEEE, 2018, pp. 1--5.

\bibitem{krestinskaya2018learning}
------, ``Learning in memristive neural network architectures using analog
  backpropagation circuits,'' \emph{IEEE Transactions on Circuits and Systems
  I: Regular Papers}, vol.~66, no.~2, pp. 719--732, 2018.

\bibitem{wang2021multi}
Y.~Wang, X.~Song, G.~Gong, and N.~Li, ``A multi-scale feature extraction-based
  normalized attention neural network for image denoising,''
  \emph{Electronics}, vol.~10, no.~3, p. 319, 2021.

\bibitem{wang2021channel}
Y.~Wang, X.~Song, and K.~Chen, ``Channel and space attention neural network for
  image denoising,'' \emph{IEEE Signal Processing Letters}, vol.~28, pp.
  424--428, 2021.

\bibitem{wei20203}
K.~Wei, Y.~Fu, and H.~Huang, ``3-d quasi-recurrent neural network for
  hyperspectral image denoising,'' \emph{IEEE transactions on neural networks
  and learning systems}, vol.~32, no.~1, pp. 363--375, 2020.

\bibitem{zheng2018error}
X.~Zheng, R.~Zarcone, D.~Paiton, J.~Sohn, W.~Wan, B.~Olshausen, and H.-S.~P.
  Wong, ``Error-resilient analog image storage and compression with
  analog-valued rram arrays: an adaptive joint source-channel coding
  approach,'' in \emph{2018 IEEE International Electron Devices Meeting
  (IEDM)}.\hskip 1em plus 0.5em minus 0.4em\relax IEEE, 2018, pp. 3--5.

\bibitem{tang2020fully}
Z.~Tang, Y.~Chen, S.~Ye, R.~Hu, H.~Wang, J.~He, Q.~Huang, and S.~Chang, ``Fully
  memristive spiking-neuron learning framework and its applications on pattern
  recognition and edge detection,'' \emph{Neurocomputing}, vol. 403, pp.
  80--87, 2020.

\bibitem{zhang2021quantized}
Y.~Zhang, Z.~Wu, S.~Liu, Z.~Guo, Q.~Chen, P.~Gao, P.~Wang, and G.~Liu, ``A
  quantized convolutional neural network implemented with memristor for image
  denoising and recognition,'' \emph{Frontiers in Neuroscience}, vol.~15, 2021.

\bibitem{slavova2019memristor}
A.~Slavova, ``Memristor cnn model for image denoising,'' in \emph{2019 26th
  IEEE International Conference on Electronics, Circuits and Systems
  (ICECS)}.\hskip 1em plus 0.5em minus 0.4em\relax IEEE, 2019, pp. 221--224.

\bibitem{tang2020multilayer}
Z.~Tang, R.~Zhu, R.~Hu, Y.~Chen, E.~Q. Wu, H.~Wang, J.~He, Q.~Huang, and
  S.~Chang, ``A multilayer neural network merging image preprocessing and
  pattern recognition by integrating diffusion and drift memristors,''
  \emph{IEEE Transactions on Cognitive and Developmental Systems}, 2020.

\bibitem{tian2000noise}
H.~Tian, ``Noise analysis in cmos image sensors,'' 2000.

\bibitem{cao2011chip}
Y.~Cao and X.~Zhang, ``An on-chip hot pixel identification and correction
  approach in cmos imagers,'' in \emph{2011 International SoC Design
  Conference}.\hskip 1em plus 0.5em minus 0.4em\relax IEEE, 2011, pp. 408--411.

\bibitem{singh2016speckle}
P.~Singh and R.~Shree, ``Speckle noise: Modelling and implementation,'' in
  \emph{published in International Journal of Control Theory and
  Applications}.\hskip 1em plus 0.5em minus 0.4em\relax International Science
  Press, 2016, vol.~9, no.~17, pp. 8717--8727.

\bibitem{wang2021taichi}
X.~Wang, R.~Pinkham, M.~A. Zidan, F.-H. Meng, M.~P. Flynn, Z.~Zhang, and W.~D.
  Lu, ``Taichi: A tiled architecture for in-memory computing and heterogeneous
  integration,'' \emph{IEEE Transactions on Circuits and Systems II: Express
  Briefs}, 2021.

\bibitem{lecun1998gradient}
Y.~LeCun, L.~Bottou, Y.~Bengio, and P.~Haffner, ``Gradient-based learning
  applied to document recognition,'' \emph{Proceedings of the IEEE}, vol.~86,
  no.~11, pp. 2278--2324, 1998.

\bibitem{krizhevsky2009learning}
A.~Krizhevsky, G.~Hinton \emph{et~al.}, ``Learning multiple layers of features
  from tiny images,'' 2009.

\bibitem{boncelet2009image}
C.~Boncelet, ``Image noise models,'' in \emph{The Essential Guide to Image
  Processing}.\hskip 1em plus 0.5em minus 0.4em\relax Elsevier, 2009, pp.
  143--167.

\bibitem{wang2004image}
Z.~Wang, A.~C. Bovik, H.~R. Sheikh, and E.~P. Simoncelli, ``Image quality
  assessment: from error visibility to structural similarity,'' \emph{IEEE
  transactions on image processing}, vol.~13, no.~4, pp. 600--612, 2004.

\bibitem{krestinskaya2019memristive}
O.~Krestinskaya, A.~Irmanova, and A.~P. James, ``Memristive non-idealities: Is
  there any practical implications for designing neural network chips?'' in
  \emph{2019 IEEE International Symposium on Circuits and Systems
  (ISCAS)}.\hskip 1em plus 0.5em minus 0.4em\relax IEEE, 2019, pp. 1--5.

\bibitem{krestinskaya2020analogue}
O.~Krestinskaya and A.~P. James, ``Analogue neuro-memristive convolutional
  dropout nets,'' \emph{Proceedings of the Royal Society A}, vol. 476, no.
  2242, p. 20200210, 2020.

\bibitem{guo2020unsupervised}
W.~Guo, M.~E. Fouda, H.~E. Yantir, A.~M. Eltawil, and K.~N. Salama,
  ``Unsupervised adaptive weight pruning for energy-efficient neuromorphic
  systems,'' \emph{Frontiers in Neuroscience}, vol.~14, p. 1189, 2020.

\bibitem{goyal2020image}
B.~Goyal, A.~Dogra, S.~Agrawal, B.~S. Sohi, and A.~Sharma, ``Image denoising
  review: From classical to state-of-the-art approaches,'' \emph{Information
  fusion}, vol.~55, pp. 220--244, 2020.

\bibitem{liu2020neural}
Z.~Liu, J.~Tang, B.~Gao, P.~Yao, X.~Li, D.~Liu, Y.~Zhou, H.~Qian, B.~Hong, and
  H.~Wu, ``Neural signal analysis with memristor arrays towards high-efficiency
  brain--machine interfaces,'' \emph{Nature communications}, vol.~11, no.~1,
  pp. 1--9, 2020.

\bibitem{cai2019fully}
F.~Cai, J.~M. Correll, S.~H. Lee, Y.~Lim, V.~Bothra, Z.~Zhang, M.~P. Flynn, and
  W.~D. Lu, ``A fully integrated reprogrammable memristor--cmos system for
  efficient multiply--accumulate operations,'' \emph{Nature Electronics},
  vol.~2, no.~7, pp. 290--299, 2019.

\bibitem{wang2019deep}
Q.~Wang, X.~Wang, S.~H. Lee, F.-H. Meng, and W.~D. Lu, ``A deep neural network
  accelerator based on tiled rram architecture,'' in \emph{2019 IEEE
  International Electron Devices Meeting (IEDM)}.\hskip 1em plus 0.5em minus
  0.4em\relax IEEE, 2019, pp. 14--4.

\bibitem{lammie2019variation}
C.~Lammie, O.~Krestinskaya, A.~James, and M.~R. Azghadi, ``Variation-aware
  binarized memristive networks,'' in \emph{2019 26th IEEE International
  Conference on Electronics, Circuits and Systems (ICECS)}.\hskip 1em plus
  0.5em minus 0.4em\relax IEEE, 2019, pp. 490--493.

\bibitem{saxena2010indirect}
V.~Saxena and R.~J. Baker, ``Indirect compensation techniques for three-stage
  fully-differential op-amps,'' in \emph{2010 53rd IEEE International Midwest
  Symposium on Circuits and Systems}.\hskip 1em plus 0.5em minus 0.4em\relax
  IEEE, 2010, pp. 588--591.

\bibitem{danesh2019ultra}
M.~Danesh, A.~Jayaraj, S.~T. Chandrasekaran, and A.~Sanyal, ``Ultra-low power
  analog multiplier based on translinear principle,'' in \emph{2019 IEEE
  International Symposium on Circuits and Systems (ISCAS)}.\hskip 1em plus
  0.5em minus 0.4em\relax IEEE, 2019, pp. 1--5.

\bibitem{khwa201865nm}
W.-S. Khwa, J.-J. Chen, J.-F. Li, X.~Si, E.-Y. Yang, X.~Sun, R.~Liu, P.-Y.
  Chen, Q.~Li, S.~Yu \emph{et~al.}, ``A 65nm 4kb algorithm-dependent
  computing-in-memory sram unit-macro with 2.3 ns and 55.8 tops/w fully
  parallel product-sum operation for binary dnn edge processors,'' in
  \emph{2018 IEEE International Solid-State Circuits Conference-(ISSCC)}.\hskip
  1em plus 0.5em minus 0.4em\relax IEEE, 2018, pp. 496--498.

\bibitem{yildirim2021analog}
M.~Yildirim, ``Analog circuit architecture for max and min pooling methods on
  image,'' \emph{Analog Integrated Circuits and Signal Processing}, vol. 108,
  no.~1, pp. 119--124, 2021.

\bibitem{wanimage}
Z.~Wan and K.~Lee, ``Image pre-processor for robust deep neural network
  inference hardware.''

\bibitem{mahmoud2017ideal}
M.~Mahmoud, B.~Zheng, A.~D. Lascorz, F.~Heide, J.~Assouline, P.~Boucher,
  E.~Onzon, and A.~Moshovos, ``Ideal: Image denoising accelerator,'' in
  \emph{2017 50th Annual IEEE/ACM International Symposium on Microarchitecture
  (MICRO)}.\hskip 1em plus 0.5em minus 0.4em\relax IEEE, 2017, pp. 82--95.

\end{thebibliography}

\end{document}